\documentclass[aps,prb,showpacs,twocolumn]{revtex4}
\usepackage{graphicx}
\usepackage[latin1]{inputenc}
\usepackage{amsmath}
\newcommand{\eq}[1]{Eq.~(\ref{#1})}
\newcommand{\fig}[1]{Fig.~\ref{#1}}

\begin{document}
\title
{Spontaneous breaking of the Fermi surface symmetry in the $t-J$
model: a
 numerical study}
\author{Bernhard Edegger$^{1,2,3}$, V.~N.~Muthukumar$^2$,
        and Claudius Gros$^1$}
\affiliation{$^1$ Institute for Theoretical Physics,
Universität Frankfurt, D-60438 Frankfurt, Germany}
\affiliation{$^2$ Department of Physics, City College of the City
University of New York, New York, NY 10031} \affiliation{$^3$
Department of Physics, Princeton University, Princeton, NJ 08544}
\date{\today}

\begin{abstract}
We present a variational Monte Carlo (VMC) study of spontaneous
Fermi surface symmetry breaking in the $t-J$ model. We find that
the variational energy of a Gutzwiller projected Fermi sea is
lowered by allowing for a finite asymmetry between the $x$- and
the $y$-directions. However, the best variational state remains a
pure superconducting state with $d$-wave symmetry, as long as the
underlying lattice is isotropic. Our VMC results are in good
overall agreement with slave boson mean field theory (SBMFT) and
renormalized mean field theory (RMFT), although apparent
discrepancies do show up in the half-filled limit, revealing some
limitations of mean field theories. VMC and complementary
RMFT calculations also confirm the SBMFT predictions that
many-body interactions can enhance any anisotropy in the underlying
crystal lattice. Thus, our results may be of consequence for the
description of strongly correlated superconductors with an
anisotropic lattice structure.
\end{abstract}
\maketitle

\section{Introduction}
The phenomenon of high temperature superconductivity has led to
intensive debates about possible superconducting states in purely
repulsive models, such as the single band Hubbard model in two
dimensions. Various strong coupling approaches show that a
repulsive onsite (Hubbard) interaction $U$ may lead to
instabilities of the two-dimensional (2D) Fermi sea, such as
antiferromagnetism, $d$-wave superconductivity \textit{etc.} An
interesting possibility is the tetragonal symmetry breaking of the
2D Fermi surface due to strong electron correlations, which may
result in a quasi one-dimensional (1D) state. This instability,
which competes with, and seems to be suppressed by the $d$-wave
pairing state, was first reported by Yamase and Kohno within slave
boson mean field theory (SBMFT) in the $t-J$ model.
\cite{Yamase00} The same effect was seen within a renormalization
group (RG) study in the Hubbard model by Halboth and Metzner who
called it the ``Pomeranchuk instability" (PI) of the Fermi
surface.\cite{Halboth00} Subsequently, several authors studied
this problem.
\cite{Valenzuela01,Grote02,Honerkamp02,Neumayr03,Yamase04,Miyanaga05,
Metzner03,Khavkine04,Yamase05,DellAnna06,Yamase06,Kao05,Carter04}
It was also argued that the tendency towards a quasi 1D state may
enhance a bare anisotropy of the underlying lattice structure, as
for example present in YBa$_2$Cu$_3$O$_{6+y}$ (YBCO).
\cite{Yamase06}

Previous studies of the PI for the $t-J$ and the Hubbard models
were mainly based on SBMFT and RG methods. Clearly, it is desirable
to use an alternative approach,
and examine if indeed a PI is present. In this paper,
we analyze this problem comprehensively using a
variational Monte Carlo
(VMC) method.\cite{Gros88,Yokoyama88,Becca00,Paramekanti01}
Our VMC results agree qualitatively with previous SBMFT
studies over a wide doping range except for discrepancies
that show up very close to half-filling. We also supplement
the VMC results by presenting results from
renormalized
mean field theory (RMFT)\cite{Zhang88,Edegger06} calculations.
These results agree well
with those from SBMFT, but reveal similar deviations from
the VMC results near half-filling. We argue that the disagreement
stems from limitations of the mean field theories when dealing
with nearly half-filled states.

The outline of the paper is as follows: In Section II, we introduce
the $t-J$ Hamiltonian and describe the VMC scheme. The VMC results
for a quasi 1D state in an isotropic $t-J$ model are presented in
Section III. We discuss the magnitude of asymmetry, the condensation
energy per site and the model parameter dependence as a function
of the hole concentration (doping). In Section IV, we provide an RMFT study for
the PI and give an explicit comparison of the Gutzwiller
renormalization scheme (GRS) with the VMC results. Section V is
dedicated to a discussion of RMFT and VMC results for the
anisotropic $t-J$ model and is followed by our conclusions.

\section{Model and VMC scheme}

We consider the $t-J$ model in two dimensions,
\begin{equation}
H=- \sum_{i,j,\sigma} {t_\tau}\, c_{i,\sigma}^\dagger
c_{j,\sigma}+ \, J \sum_{\langle i,j \rangle} \left( {\bf S}_i \,
{\bf S}_j - \frac 1 4 n_i n_j \right)\, ,
\end{equation}
defined in a projected Hilbert space where double occupancies are
forbidden. In the above equation, ${\bf S}_i$ and $n_i$ are
respectively, electron spin and density operators at site $i$. The
kinetic energy is determined by $t_\tau$, the $\tau$th neighbor
hopping integral. Throughout this paper, we denote
nearest neighbor (n.n.)  hopping terms by
$t$, and choose next n.n. hopping terms $t'=0$ or $t'=-0.3t$. For
simplicity, we only consider the superexchange coupling $J = 4t^2/U$
between n.n. sites, and choose the value, $J=0.3t$.

We implement the no-double occupancy constraint in the $t-J$
Hamiltonian by working with Gutzwiller projected states $|\Psi
\rangle=P_G\,|\Psi_0\rangle$, where the Gutzwiller projection
operator, $P_G=\prod_i (1-n_{i\uparrow}n_{i\downarrow})$. Such
wavefunctions were initially proposed as variational states to
describe superconductivity in the proximity of a Mott insulating
phase. \cite{Anderson87,Zhang88,Gros88,Yokoyama88} The VMC
technique which allows for a numerical evaluation of expectation
values in these states successfully predicted $d-$wave pairing in
the $t-J$ model. \cite{Gros88,Yokoyama88} The method has been
extended more recently to study the coexistence of (and
competition between) various phases like antiferromagnetism, flux
states, and superconductivity in the $t-J$ model
\cite{Chen90,Himeda99,Shih04,Ivanov04}. Motivated by
the phenomenology of the high temperature superconductors, other
improvements such as the inclusion of longer range hopping terms
\cite{Shih05}, increase of the number of variational parameters,
and introduction of long range correlations through Jastrow
factors \cite{Sorella02} have also been proposed. Here, we extend
previous works by allowing a symmetry breaking between the $x-$
and the $y-$direction in the variational wavefunction $|\Psi
\rangle$.

The ground state wave function is written in
the form,
\begin{equation}
|\Psi \rangle\,=\,P_G |\Psi_{\rm BCS}\rangle \ ,
\end{equation}
where $|\Psi_{BCS}\rangle=\left(\sum_k \, a_k  \, c^\dagger_{k\uparrow}
c^\dagger_{-k\downarrow}\right)^{N/2} | 0 \rangle $ is the $N$-electron
BCS wave function with
\begin{equation}
a_k\,\equiv\,\frac {v_k}{u_k} \,=
\frac{\Delta_k}{\zeta_k+\sqrt{\zeta_k^2+\Delta_k^2}}\ .
\end{equation}
To allow for a possible quasi 1D state as well as
for finite $d$- and $s$-wave pairing, we choose,
\begin{eqnarray}
\zeta_k=&-&2\,[(1+\delta_{\rm var}^{\rm 1D}) \cos k_x +
(1-\delta_{\rm
var}^{\rm 1D})\cos k_y] \nonumber \\
&-& 4 t'_{\rm var} \cos k_x \cos k_y\,-\, \mu_{\rm var}
\end{eqnarray}
and
\begin{equation}
\Delta_k=\Delta_{\rm var}^{(d)}(\cos {k_x}-\cos{k_y})+ \Delta_{\rm
var}^{(s)}(\cos {k_x}+\cos {k_y})\ .
\end{equation}
We then have the following five variational parameters: (i) the
asymmetry $\delta^{\rm 1D}_{\rm var}$ between n.n.\ $x-$ and
$y-$ hopping matrix elements;
(ii) the variational next n.n.\ hopping term $t'_{\rm
var}$; (iii) a variational chemical potential $\mu_{\rm var}$;
(iv) and (v) variational parameters for $d-$ and $s-$wave pairing,
$\Delta_{\rm var}^{(d)}$ and $\Delta_{\rm var}^{(s)}$,
respectively. Using standard VMC techniques \cite{Gros89}, we
compute energy expectation values and minimize the energy by
searching for the optimal set of variational parameters. We use
tilted lattices with periodic boundary conditions. Typically,
$10^5$ Monte Carlo steps are performed for each set of variational
parameters. The resulting statistical errors are given in the
relevant figures by error bars.

\section{VMC results for the isotropic $t-J$ model}

\begin{figure}
  \centering
 \includegraphics*[width=0.5\textwidth]{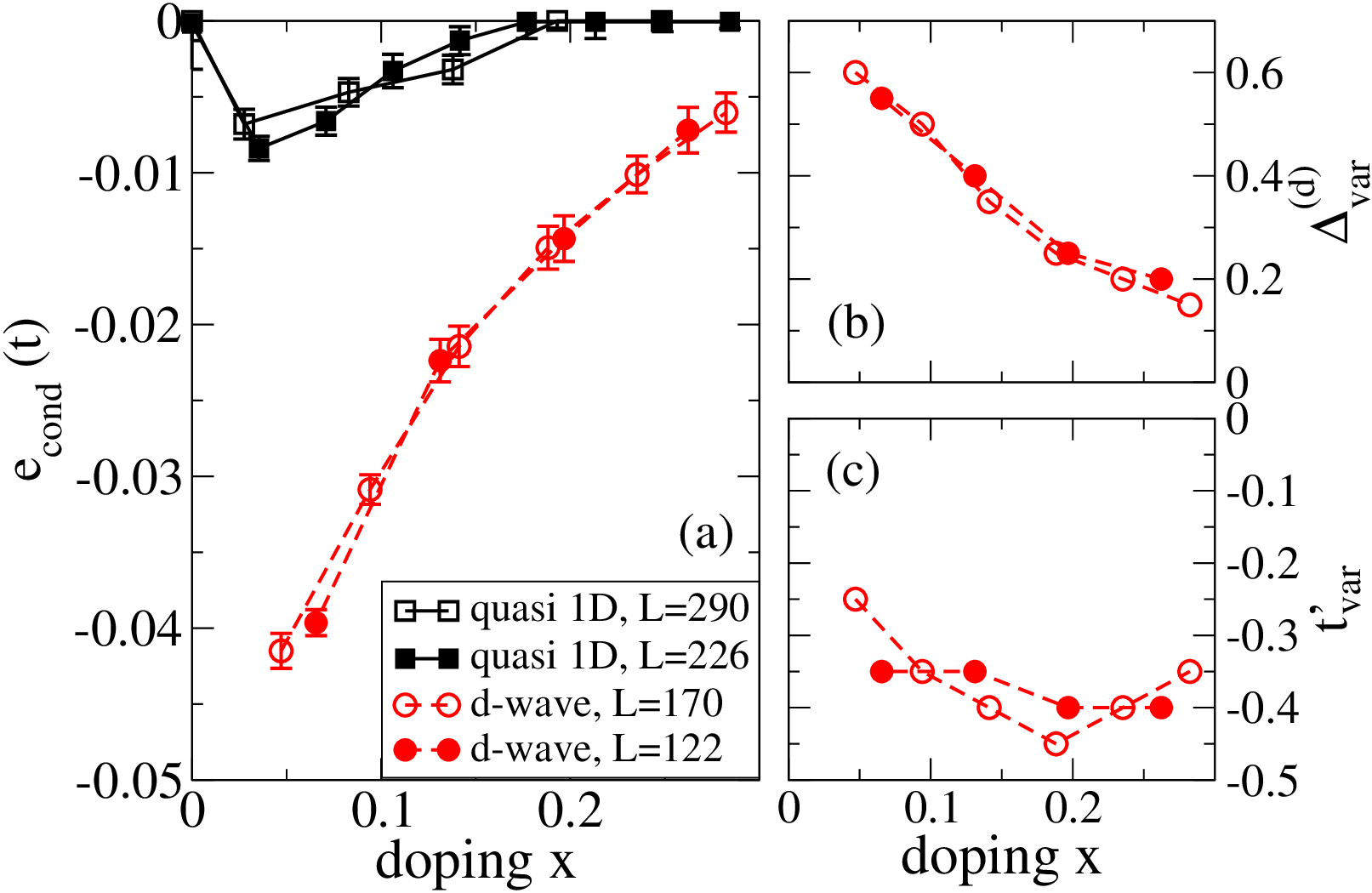}
 \caption{(color online)
 (a) VMC results for condensation energies per site $e_{\rm cond}$
 of the quasi 1D state ($\Delta_k\equiv 0$) and
 the $d$-wave state ($\Delta_k \neq 0$) with $t'=-0.3\,t$.
 Optimal variational parameters $\Delta^{(d)}_{\rm var}$ and $t'_{\rm var}$
 of the $d$-wave state are shown in (b) and (c). The errors
 in (b) and (c) are $\Delta \Delta^{(d)}_{\rm var} =0.05$ and $\Delta
t'_{\rm var}=0.05$, respectively. System sizes: $L=11^2+1=122$,
$L=13^2+1=170$, $L=15^2+1=226$, and $L=17^2+1=290$.
 }
 \label{compareDwave}
\end{figure}

\begin{figure}
\centering
 \includegraphics[width=0.45\textwidth]{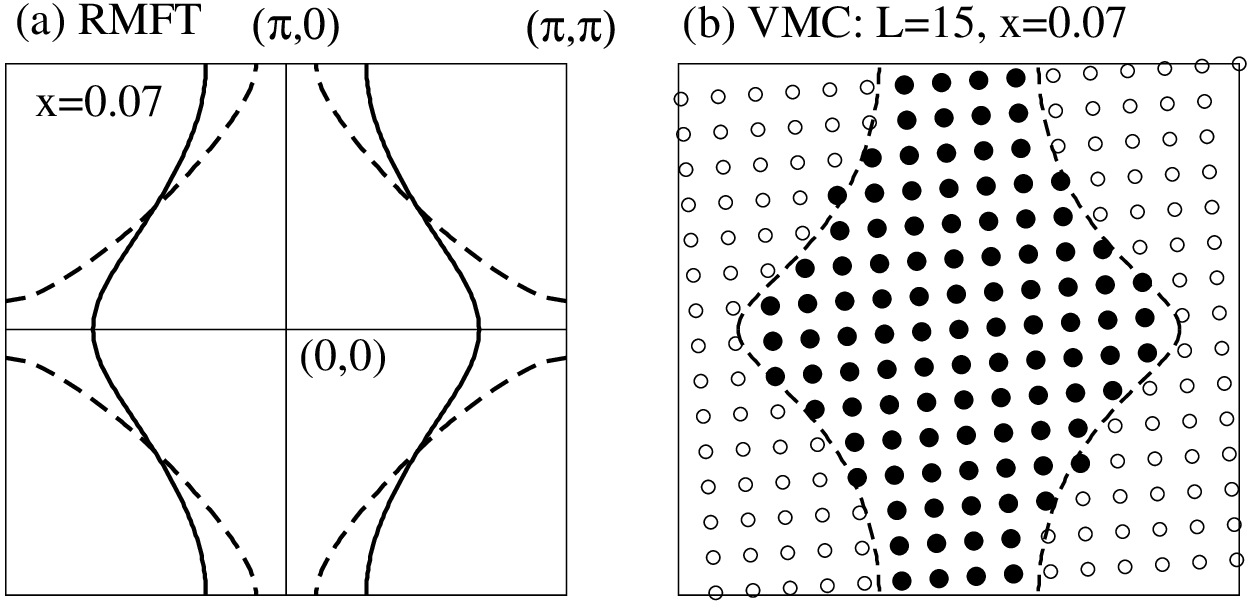}
 \caption{Fermi surface of the isotropic $t-J$ model
 with $J=0.3t$ and $t'=-0.3t$
 at $x=0.07$ (a) RMFT results for the Fermi surface
 of the normal state with $\Delta_k\equiv0$
 (solid line) and the
 optimal $d$-wave state (dashed line).
 (b) Best quasi 1D state on a $(15^2+1)$-sites lattice by VMC;
 filled circles indicate
 the Fermi surface.}
 \label{FS}
\end{figure}

\begin{figure}
  \centering
 \includegraphics*[width=0.5\textwidth]{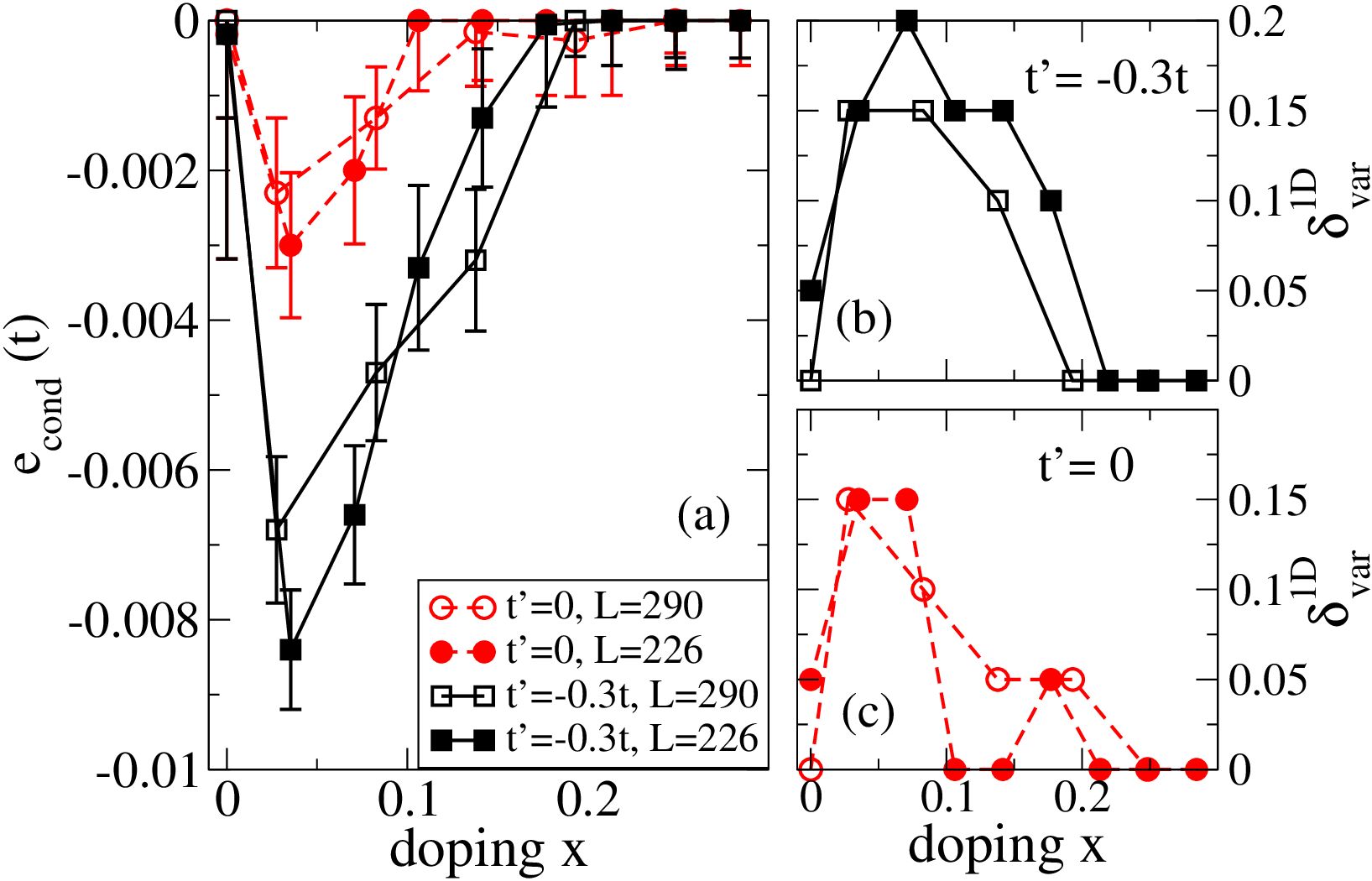}
 \caption{(color online) VMC results for a quasi 1D state
 for the isotropic $t-J$ model ($J=0.3 t$) with
 $t'=0$  (circles) and $t'=-0.3 t$ (squares). Doping dependence $x$
 of (a) the condensation energy per site, $e_{\rm cond}$, and
  (b),(c)
 the optimal variational $\delta^{\rm 1D}_{\rm var}$.
 The errors in (b) and (c) are $\Delta \delta^{\rm 1D}_{\rm var} =0.05$.
 System sizes: $L=15^2+1=226$ and $L=17^2+1=290$.
 }\label{comparePom}
\end{figure}

\begin{figure}
  \centering
 \includegraphics*[width=0.34\textwidth]{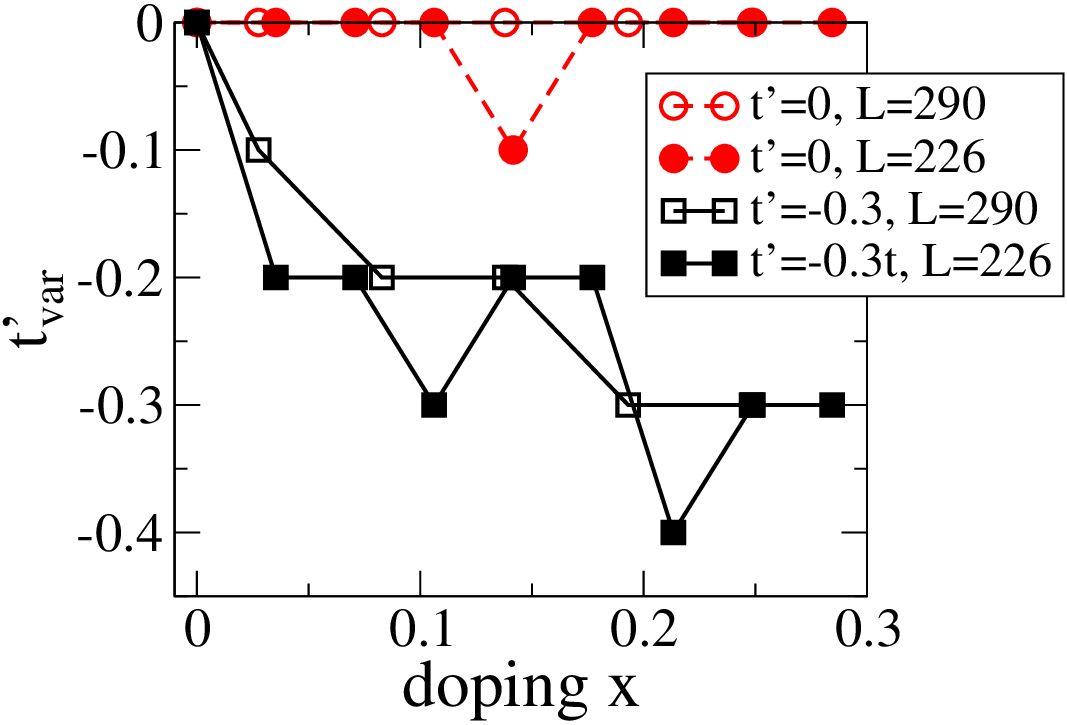}
 \caption{(color online) VMC results for the quasi 1D state
 in the isotropic $t-J$ model ($J=0.3 t$) with
$t'=0$ (circles) and $t'=-0.3 t$ (squares). Doping dependence $x$
 of the optimal variational $t'_{\rm var}$. $\Delta
 t'_{\rm var}=0.1$; System sizes: $L=15^2+1=226$ and
 $L=17^2+1=290$.
 }\label{varHop}
\end{figure}

We first present results for
the isotropic $t-J$ Hamiltonian, with model parameters $J=0.3t$ and $t'=-0.3t$.
These are reasonable model parameters for the phenomenology of the high temperature superconductors.
The optimal solution for various values of hole concentration, $x=0-0.3$, is determined by
searching in the whole variational parameter space. We find that the pure
isotropic projected $d-$wave state always optimizes the ground
state energy, \textit{i.e.}, the $s-$wave parameter $\Delta_{\rm
var}^{(s)}$ and the asymmetry $\delta^{\rm 1D}_{\rm var}$ vanish
for all values of $x$, within our numerical resolution
[$\Delta\Delta_{\rm var}^{(s)}=0.05$ and $\Delta\delta^{\rm
1D}_{\rm var}=0.05$].

In \fig{compareDwave}(a) (circles) we show the condensation energy
per site, $e_{\rm cond}$, of the optimal state with respect to the
projected isotropic Fermi sea. The condensation energy, $e_{\rm
cond}$, is calculated by comparing the VMC energy expectation
values in the projected Fermi sea and optimal $d-$wave states. We
see a continuous increase of $|e_{\rm cond}|$ and of the
superconducting $d-$wave parameter [shown in
\fig{compareDwave}(b)] as doping $x$ decreases. The optimal
variational value for $t'_{\rm var}$ is given in
\fig{compareDwave}(c). These results show that the additional
variational parameters $\Delta_{\rm var}^{(s)}$ and $\delta^{\rm
1D}_{\rm var}$ are not relevant for improving the ground state in
the isotropic $t-J$ model.

To uncover the PI, it is necessary to suppress superconductivity
by setting $\Delta_k \equiv 0$. Doing so, our VMC calculations indeed
reveal a PI; we obtain an improvement of the
energy expectation value relative to the isotropic projected
Fermi sea by using a finite asymmetry $\delta^{\rm 1D}_{\rm var}$,
although the underlying lattice is still isotropic. We compare the
condensation energy of this state to that of the $d-$wave state.
Results are shown in \fig{compareDwave}(a). As for the $d-$wave
state, $|e_{\rm cond}|$ for the quasi 1D state initially increases
as $x$ decreases. However its energy gain is much less than that
of the $d-$wave, and so the latter is always favored on an
isotropic lattice. Furthermore, note that the condensation of the
quasi 1D state saturates and finally vanishes very close to
half-filling. We will come back to this point later.

The resulting VMC Fermi surface of the optimal quasi 1D state at
$x=0.07$ is shown in \fig{FS}(b). It reveals why finite size
effects become important in the VMC calculations when dealing with
a finite asymmetry in the projected Fermi sea. Varying
$\delta^{\rm 1D}_{\rm var}$ causes discontinuous changes of the
Fermi surface on a finite lattice. The occupied states
regroup for certain $\delta^{\rm 1D}_{\rm var}$ leading to
small yet discontinuous changes in the FS as a function of the
variational parameter $\delta^{\rm 1D}_{\rm var}$.

These finite size effects cause a rather large error for the
optimal value of asymmetry, $\Delta \delta^{\rm 1D}_{\rm var}
=0.05$, and for the effective next n.n.\ hopping, $\Delta t'_{\rm
var}=0.05-0.1$.  The jump size increases with decreasing system
size, thus requiring sufficiently
large lattices. The problem is less severe when considering a
superconducting state, where the occupancy in momentum space
changes continuously at the Fermi surface.

To consider the effect of $t'$ on the PI, we compared the two cases,
$t'=0$ and $t'=-0.3t$, in the absence of superconducting
order ($\Delta_k\equiv0$). The PI is stronger for $t'<0$,
yielding a larger condensation energy [\fig{comparePom}(a)].
As seen in \fig{comparePom}(b), the anisotropy is significant
even at higher doping levels and exists
up to $x\approx0.2$  for $t'=-0.3t$. For $t'=0$, the $\delta^{\rm
1D}_{\rm var}$ is significant only in the range, $x=0.03-0.10$ [\fig{comparePom}(c)].
In \fig{varHop} we show the optimal variational value for $t'_{\rm
var}$. Interestingly, $t'_{\rm var} \to 0$ for
$x\to0$ even for a bare dispersion $t'=-0.3t$. Recently,
we reported a similar renormalization of
the next n.n.\ hopping terms due to strong coupling effects
within RMFT.
\cite{Edegger06,Gros06}

Although there is good overall agreement between our
VMC data and SBMFT results, we find
clear and significant discrepancies in the limit $x\to 0$. As
seen in \fig{comparePom}(a)-(c) the asymmetry goes to zero at
$x=0$ within our VMC calculations. On the other hand, SBMFT as well
as RMFT (discussed in the next section) predict a pure 1D state at half-filling
when $\Delta_k\equiv0$. This hints at limitations of the mean field
theories when treating states near half-filling. We shall
discuss this in more detail, after discussing results from the RMFT calculation
for the PI.

\section{RMFT and Gutzwiller renormalization for the isotropic $t-J$ model}

\begin{figure}
  \centering
 \includegraphics*[width=0.5\textwidth]{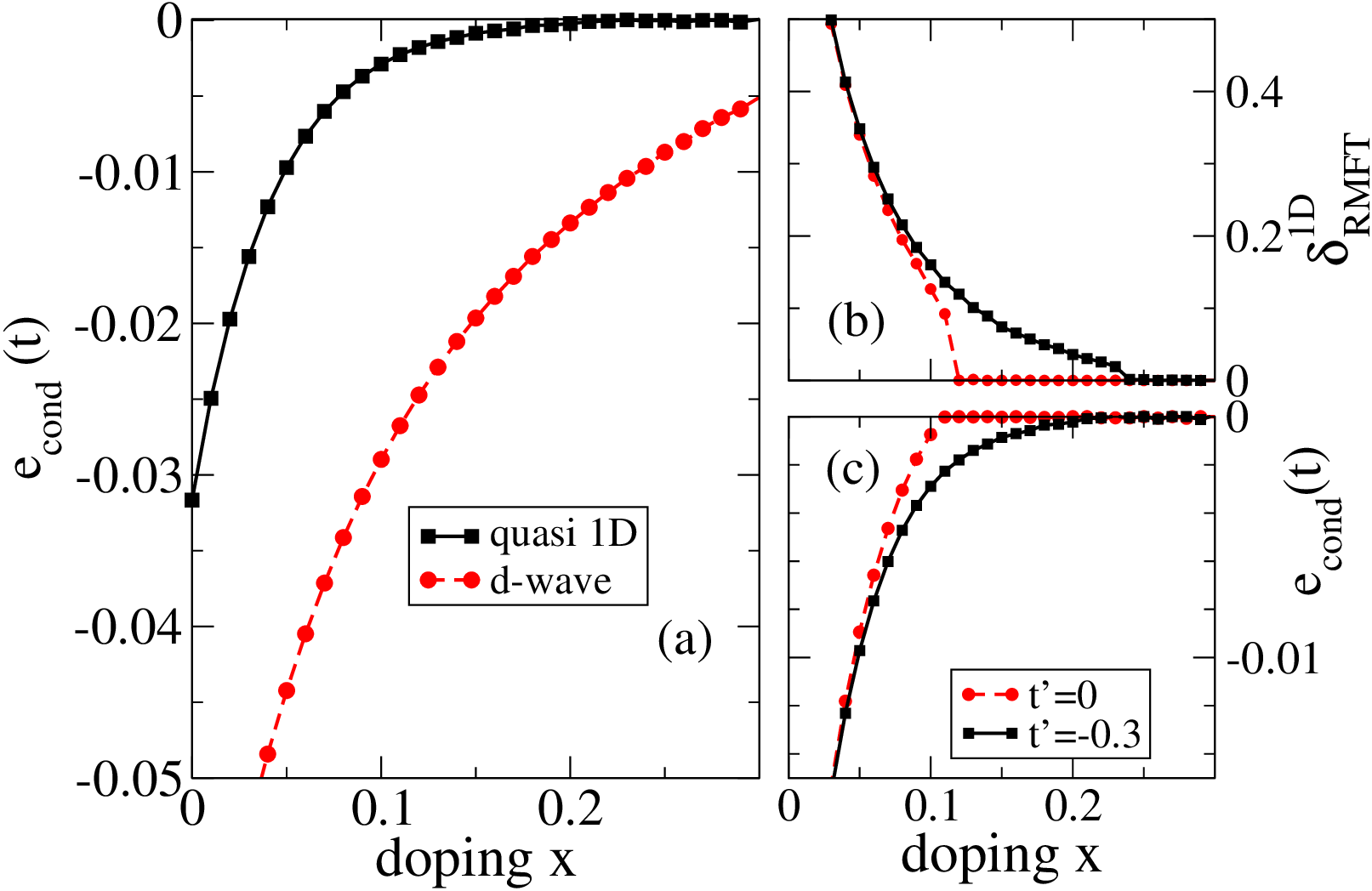}
 \caption{(color online) RMFT results for the isotropic $t-J$ model with $J=0.3 t$.
 (a) Condensation
 energy per site relative to the isotropic non-SC solution ($\Delta_k\equiv0$, $\xi_x=\xi_y$)
 for the quasi 1D state ($\Delta_\tau\equiv0$) and for the optimal $d$-wave
 state with $t'=-0.3 t$. (b) Asymmetry and (c) condensation
 energy per site for the quasi 1D state with $t'=0$ and $t'=-0.3$,
 respectively.
 }\label{RMFTPom}
\end{figure}

In this section, we present some results from renormalized mean
field theory (RMFT). The RMFT gap equations have been derived in
previous works \cite{Zhang88,Edegger06}. Here, we solve them
allowing for possible anisotropic solutions. We find that the
optimal self consistent state remains a pure $d$-wave
superconductor. This is consistent with the VMC data presented in
the previous section, and with SBMFT calculations.\cite{Yamase00}
The condensation energy of this optimal state relative to the
isotropic renormalized Fermi sea is shown in \fig{RMFTPom}. To
uncover the PI, we now look for solutions with the constraint
$\Delta_k\equiv0$, as before. Doing so, we find two self
consistent solutions for sufficiently small doping: an isotropic
and an anisotropic renormalized Fermi sea. We compare the
condensation energy of the anisotropic quasi 1D state with that of
the $d$-wave pairing state for a bare dispersion $t'=-0.3t$. As
seen from \fig{RMFTPom}(a), the doping dependence as well as the
magnitude of the condensation energy agree well with the VMC
results in \fig{compareDwave}(a). However, the two results begin
to differ in the vicinity of $x=0$, where the condensation energy
of the quasi 1D state vanishes in the VMC scheme.

Within RMFT, the order parameter characterizing the asymmetry,
$\delta^{\rm
1D}_{\rm RMFT}$ is given by,
\begin{equation}
\delta^{\rm 1D}_{\rm RMFT}\, \equiv\, \frac{\tilde t_x - \tilde
t_y}{\tilde t_x+\tilde t_y}\ .
\end{equation}
Here, the effective hopping $\tilde t_\tau$ in the
$\tau$-direction is related to the bare hopping $t_\tau$
by\cite{hoppings},
\begin{equation}
\tilde t_{\tau}\,\equiv\,g_t\,t_\tau\,+\,\frac{3
g_s-1}8\,J\,\xi_\tau \ , \label{eq_hop}
\end{equation}
where $J=4 t^2_\tau/U$, $\xi_\tau \equiv \sum_\sigma \langle
c^\dagger_{i,\sigma} c_{i+\tau,\sigma} \rangle_{|\Psi_0\rangle}$,
and $\tau=x,y$. The renormalization factors for the kinetic
energy, $g_t=2x/(1+x)$,  and for the spin-spin correlation,
$g_s=4/(1+x)^2$, are derived within the Gutzwiller
Renormalization Scheme (GRS).

The origin of the PI can be seen directly from
\eq{eq_hop}. It shows that the tendency to a quasi
1D state stems from the $J$-term [2nd term in \eq{eq_hop}] because
it includes the factor $\xi_\tau$. For an isotropic dispersion
it is the only quantity in \eq{eq_hop} that may cause an anisotropy
in the effective hopping $\tilde t_\tau$. Similar arguments apply
for the enhancement of a bare asymmetry $\delta^{\rm 1D}_0$ in a
slightly anisotropic lattice (see also discussion below). The
origin of the PI may also be understood in the framework of a
Landau-Ginzburg analysis as shown by Yamase and
Kohno within SBMFT.\cite{Yamase00}

In \fig{RMFTPom}(b) and (c), we show our results for
$\delta^{\rm 1D}_{\rm RMFT}$
and the condensation energy relative to the projected (symmetric)
Fermi sea. Results are shown as a function of $x$ for two values of $t'$,
$t'=0$ and $t'=-0.3t$. We see that a negative $t'$ favors the PI,
in agreement with SBMFT and VMC results. The asymmetry
$\delta^{\rm 1D}_{\rm RMFT}$ is shown in \fig{RMFTPom}(b) and
agrees quantitatively with the VMC results in
\fig{comparePom}(b),(c) for $x>0.07$. However, for $x \to
0$ the asymmetry increases strongly within RMFT, whereas it
saturates and finally disappears within the VMC scheme.

The Fermi surface of the quasi 1D state from RMFT at $x=0.07$ is shown in
\fig{FS}(a) and agrees well with that obtained from VMC calculations for the
same doping [\fig{FS}(b)].

\begin{figure}
  \centering
  \includegraphics*[width=0.44\textwidth]{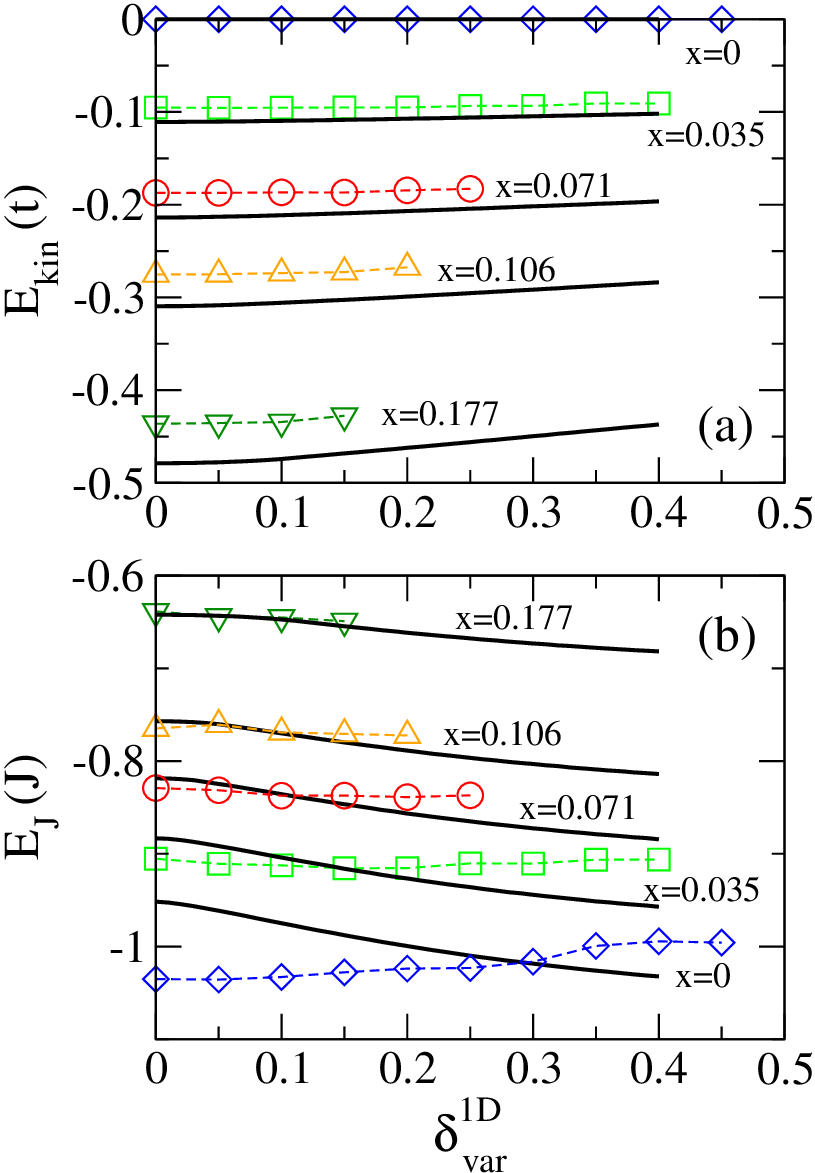}
 \caption{(color online) Comparison of the Gutzwiller renormalization scheme
 (GRS, solid lines) and VMC results (dashed lines with symbols) for (a)
  the kinetic and (b) the superexchange energy of the projected Fermi Sea
 ($\Delta_k \equiv 0$) at different doping
  levels $x$. The dependence on the asymmetry $\delta^{1D}_{\rm var}$ is illustrated.
  VMC results are taken from the $L=15^2+1$-system and
  $t'=0$. Statistical VMC errors are much smaller than the symbol size.
 }\label{compare}
\end{figure}

We now turn our attention to the discrepancy between VMC and RMFT
(and SBMFT) results near half-filling. In \fig{compare}, we plot
the kinetic energy $E_{\rm kin}$ and the superexchange energy
$E_J$ ($J$-term in the Hamiltonian) for the projected Fermi Sea
($\Delta_k \equiv 0$) as a function of the asymmetry
$\delta^{1D}_{\rm var}$ for various doping concentrations. We
compare the VMC data with the results from the Gutzwiller
renormalization scheme (GRS) for $t'=0$. \fig{compare} shows that
the GRS only approximately agrees with the nearly exact VMC
results. In particular, the behavior of the superexchange energy
as a function of $\delta^{\rm 1D}_{\rm var}$ is qualitatively
different in the two schemes. The discrepancies are acute near
half filling (see data for $x=0$ and $x=0.035$). A reasonable
agreement is obtained for larger doping. In general, the VMC data
shows a weaker dependence of the energy on the asymmetry
$\delta^{\rm 1D}_{\rm var}$.

Our VMC results are also consistent with previous studies of the
Gutzwiller wavefunction  at half-filling in the 1D limit
\cite{Gros87}, which corresponds to $\delta^{\rm 1D}_{\rm var}
\equiv 1$ in our calculations. The superexchange energy (deduced
from the n.n. spin-spin correlations) of the pure 1D state on an
isotropic 2D lattice is much worse than that of the projected
isotropic 2D Fermi sea. On the contrary a simple evaluation by the
GRS would favor a pure 1D state at half-filling. We thus
think that the drawbacks of the mean field theories partially
stem from invoking 1D physics by allowing a finite asymmetry.
We further note that the renormalization effects become
largest near half-filling. Therefore, it is naturally that
discrepancies can show up in this limit, as already seen
in previous VMC studies, e.g., for the quasiparticle weight
renormalization\cite{Fukushima05,Nave05,Yang06,Chou06}.
With that in mind, our VMC results are not really surprising.
Though the GRS gives valuable
insights into the behavior of strongly correlated electronic
systems, the above results show that a verification of GRS
results by the VMC technique is often very important. These limitations of
the GRS directly impact the RMFT (which is based on the GRS)
and the SBMFT (which is closely related to RMFT).

\section{RMFT and VMC calculations
for the anisotropic $t-J$ model}

\begin{figure}
  \centering
 \includegraphics*[width=0.5\textwidth]{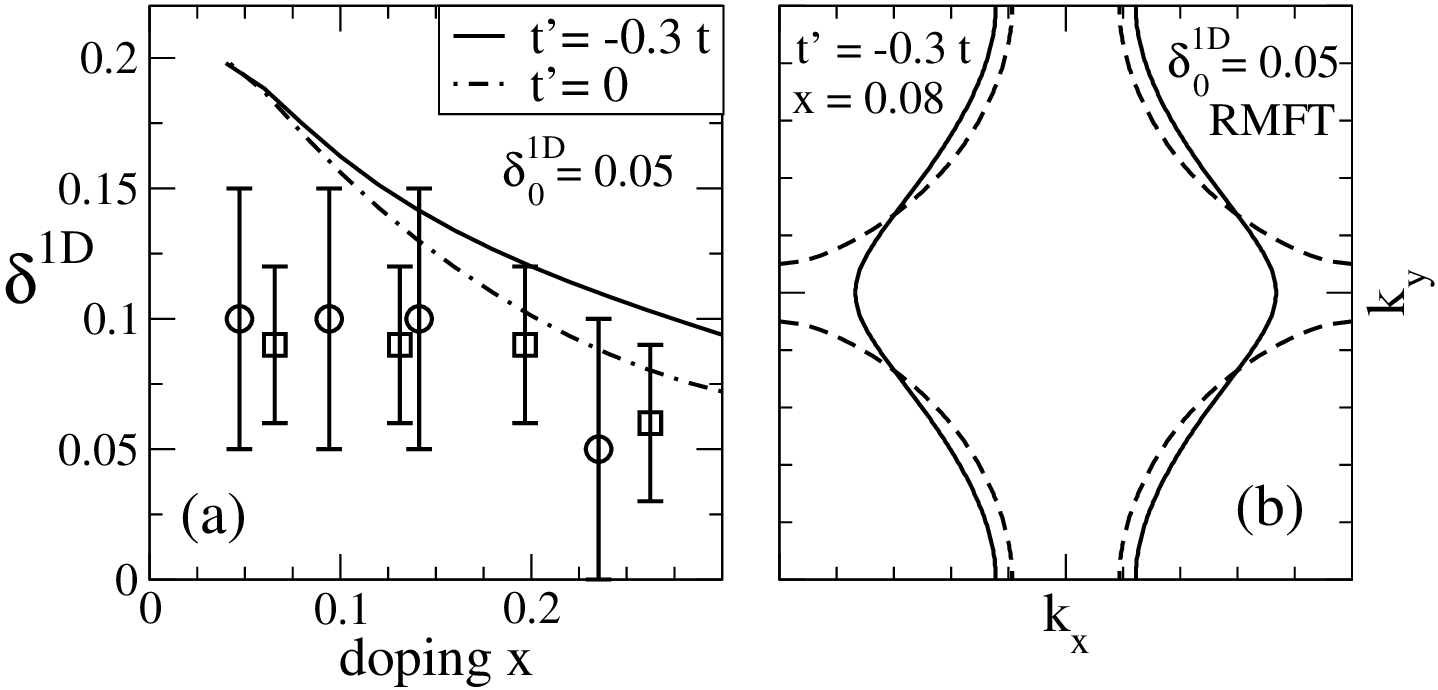}
 \caption{RMFT and VMC results for the $d+s$-wave ground state
 of the anisotropic $t-J$ model with $J=0.3 t$ and $\delta_0^{\rm 1D}\equiv (t_x-t_y)/(t_x+t_y)=0.05$.
 (a) Effective asymmetry $\delta^{\rm 1D}\equiv(\tilde t_x-\tilde t_y)/(\tilde
 t_x+\tilde t_y)$ from RMFT
 as a function of hole doping $x$ for (dashed) $t'=0$ and (solid)
 $t'=-0.3t$. VMC results for $t'=0$ are given by squares
 and circles for $L=122$ and
 $L=170$, respectively.
 (b) RMFT Fermi surface (solid lines) of the $d+s$-wave ground
 state and the tight binding dispersion (dashed) at $x=0.08$ with $t'=-0.3t$ and $\delta^0_{\rm 1D}=0.025$.
 }\label{anisotrop}
\end{figure}

Our results from VMC and RMFT confirm that a quasi 1D state is
always suppressed by the $d$-wave pairing state. A PI occurs only
when $\Delta_k\equiv0$. However, the situation can be quite
different when the underlying lattice structure is anisotropic.
In this case, the tendency towards a quasi 1D state is present
even in the superconducting state. SBMFT\cite{Yamase00}
predicts an optimal state which has a dominant $d$-wave symmetry with a
small $s$-wave contribution.\cite{smallGap} Interestingly, the bare
anisotropy $\delta^{\rm 1D}_0$ of the lattice is enhanced due to
the electron correlations. Here we re-examine this prediction
within the RMFT and VMC schemes. Results from RMFT are shown in
\fig{anisotrop}(a) and (b) and agree quantitatively with the SBMFT
data. As seen in \fig{anisotrop}(a), the bare asymmetry of
$\delta^{\rm 1D}_0=0.05$ increases within the RMFT calculations up
to about $\delta^{\rm 1D}_{\rm opt}=0.2$ in the underdoped regime.
These results are confirmed to some extent by VMC calculations for
$t'=0$ in \fig{anisotrop}(a) (circles and squares), that show an
increase of the asymmetry up to about $\delta^{\rm 1D}_{\rm
var}\approx 0.1$. However, owing to numerical difficulties,
the errors in these VMC calculations are
quite large. \cite{numerics} In
\fig{anisotrop}(b), we compare the Fermi surface obtained from the
bare dispersion ($\delta^{\rm 1D}_0=0.05$) with that of the
optimal superconducting state obtained by solving the
RMFT equations self consistently, for $x=0.08$. As seen
in the figure, the enhancement of anisotropy due to strong
correlations may even lead to a change in the topology of
the underlying Fermi surface.

\section{Conclusions}

In this paper, we performed a Variational Monte Carlo (VMC) study
of the Pomeranchuk instability in the isotropic $t-J$ model. We also
supplemented the study with results from renormalized
mean field theory (RMFT). Our results are in agreement with
earlier calculations from slave boson mean field theory (SBMFT)
and exact diagonaliziation on small clusters\cite{Miyanaga05}
and show that the instability is uncovered when the (optimal)
$d$-wave superconducting state
is suppressed. The Pomeranchuk instability is
seen for values of hole concentration
between $x\approx0.03$ and $x\approx 0.20$, depending on the
magnitude of the next
n.n.\ hopping integral $t'$. We find that a small bare
asymmetry $\delta^{\rm 1D}_0$ is enhanced within an anisotropic
$t-J$ model even in the superconducting state. Although there is good
overall agreement between the VMC calculations and results from
RMFT and SBMFT, discrepancies arise
very close to half-filling, showing the limitations
of these mean field theories in this regime.
This underlines the importance of complementary VMC
studies to confirm or disprove the results from RMFT or SBMFT calculations.

The tendency towards a quasi one dimensional state in a strongly
correlated electron systems is mainly governed by the
superexchange $J$. Since $J\propto 4t^2/U$, a small asymmetry in the
bare hopping integral $t$ becomes twice as large in the
superexchange energy. Hence, it is natural that the effects
discussed in this paper are largest in the underdoped regime,
where the dispersion is mainly determined by $J$. The tendency
towards a quasi one dimensional state may be also enhanced if
phonons are coupled to the lattice. The influence of such dynamic
effects should be clarified in future studies.

\acknowledgments B.E. thanks the Research Foundation of the City University
of New York for supporting his stay at Princeton.


\begin{thebibliography}{99}
\bibitem{hoppings} The effective hopping can be read from the RMFT
dispersion, e.g., from $\zeta_k$ in Eq.1 of
Ref.~\onlinecite{Edegger06}.
\bibitem{smallGap} For a bare asymmetry of $\delta_0^{\rm 1D}=0.05$,
the $s$-wave gap is only few percent of the $d$-wave gap.
\bibitem{numerics} To reduce the error in the asymmetry, $\delta_{\rm var}^{1D}$,
one needs a high number of very accurate calculations on
sufficiently large lattices. These requirements limit our
accurancy to $\delta_{\rm var}^{1D}=0.03-0.05$.
\end{thebibliography}
\end{document}